\documentclass[aps,pra,showpacs,twocolumn]{revtex4-1}
\usepackage{amssymb}
\usepackage{amsmath}
\usepackage{graphicx}
\usepackage{epsfig}
\usepackage{subfigure}
\usepackage{amsfonts}
\usepackage{color}
\usepackage{bm}

\begin{document}

\title{Robust unidirectional phantom helix states in the XXZ Heisenberg
model with Dzyaloshinskii-Moriya interaction}
\author{Y. B. Shi}
\author{Z. Song}
\email{songtc@nankai.edu.cn}

\begin{abstract}
The phantom helix states are a special set of degenerate eigenstates of the
XXZ Heisenberg model, which lie in the energy levels around zero energy and
are bidirectionally equal. In this work, we study the helix state in the XXZ
Heisenberg model with the Dzyaloshinskii-Moriya interaction (DMI). We show
exactly that only the helix states\ in one direction remain unchanged in the
presence of resonant DMI. Based on the Holstein--Primakoff (HP)
transformation, the quantum spin model is mapped to a boson model, which
allows us to understand the underlying mechanism. Furthermore, it also
indicates that such phantom states can be separated from the spectrum by the
strong DMI to enhance the robustness of the states. We demonstrate the
dynamic formation processes of unidirectional phantom helix states by
numerical simulations. The results indicate that the DMI as expected acts as
a filter with high efficiency.
\end{abstract}

\affiliation{School of Physics, Nankai University, Tianjin 300071, China }
\maketitle

\section{Introduction}
\label{Introduction}
Recent advances in experimental capability \cite%
{kinoshita2006quantum,trotzky2012probing,gring2012relaxation,schreiber2015observation,smith2016many,kaufman2016quantum}
have stimulated the study of\ the nonequilibrium dynamics of quantum
many-body systems, which has emerged as a fundamental and attractive topic
in condensed-matter physics. Unlike traditional protocols based on the
cooling mechanism, nonequilibrium many-body dynamics provides an alternative
way to access a new exotic quantum state with energy far from the ground
state \cite{Choi,Else,Khemani,Lindner,Kaneko,Tindall,YXMPRA,zhang2022steady}%
. On the other hand, as an excellent test bed for quantum simulators in
experiments \cite%
{zhang2017observation,bernien2017probing,barends2015digital,davis2020protecting,signoles2021glassy,trotzky2008time,gross2017quantum}%
, the atomic system allows us to verify the theoretical predictions on the
dynamics of quantum spin systems, which not only capture the properties of
many artificial systems, but also provide tractable theoretical examples for
understanding fundamental concepts in physics.

A well-known example is the spin-1/2 XXZ Heisenberg chain, which is
integrable and therefore most of its properties may be obtained exactly \cite%
{Mikeska2004one}. Recently, the discovery of highly excited many-body
eigenstates of the Heisenberg model, referred to as Bethe phantom states,
has received much attention from both theoretical \cite%
{popkov2016obtaining,popkov2017solution,popkov2020exact,popkov2021phantom,MESPRB}
and experimental approaches \cite%
{jepsen2020spin,jepsen2021transverse,hild2014far,jepsen2022long}.

In this article, we systematically elucidate the unidirectional phantom
helix states in the quantum spin XXZ\ Heisenberg model with the
Dzyaloshinskii-Moriya interaction (DMI). The DMI is an antisymmetric
exchange interaction that appears in inversion asymmetric structures and
favors perpendicular alignment of neighboring spins in a magnetic material 
\cite{dzyaloshinsky1958thermodynamic,bode2007chiral,roessler2006spontaneous}%
. In such a system, we show that only the helix states in one direction, and
the corresponding eigenenergy remains unchanged. To understand the
underlying mechanism, we map the spin model to a boson model based on the
Holstein--Primakoff (HP) transformation \cite{HTAPS,KAAPS}. It is distinctly
indicated that the DMI can not only support the unidirectional phantom helix
states but also separate them from nearby energy levels around zero energy.
We also investigate the dynamics driven by a local non-Hermitian
perturbation term. Based on the analysis of the\textbf{\ }perturbation
method, it is shown that a steady helix state emerges from some easily
prepared initial states.\ It relates to an exclusive concept in a
non-Hermitian system, the exceptional point (EP), which has no counterpart
in a Hermitian system. The EP in a non-Hermitian system occurs when
eigenstates coalesce \cite{bender2007making, moiseyev2011non,
krasnok2019anomalies} and is usually associated with the non-Hermitian phase
transition \cite{feng2013experimental, gupta2019parity}. We demonstrate the
dynamic formation processes of unidirectional phantom helix states by
numerical simulations. The results indicate that the DMI as expected acts as
a filter with high efficiency.

The rest of this paper is organized as follows: In Sec.~\ref{Unidirectional
helix states}, we introduce the model Hamiltonian and the corresponding
unidirectional phantom helix states. Based on the HP transformation, in Sec.~%
\ref{Bosonic representation} we express this Hamiltonian in a boson model
and investigate the underlying mechanism. Based on these results, the
dynamics behavior driven by a local non-Hermitian perturbation is proposed
in Sec.~\ref{Dynamic demonstration}. Sec.~\ref{Summary} concludes this paper.

\begin{figure*}[t]
\centering\includegraphics[width=0.9\textwidth]{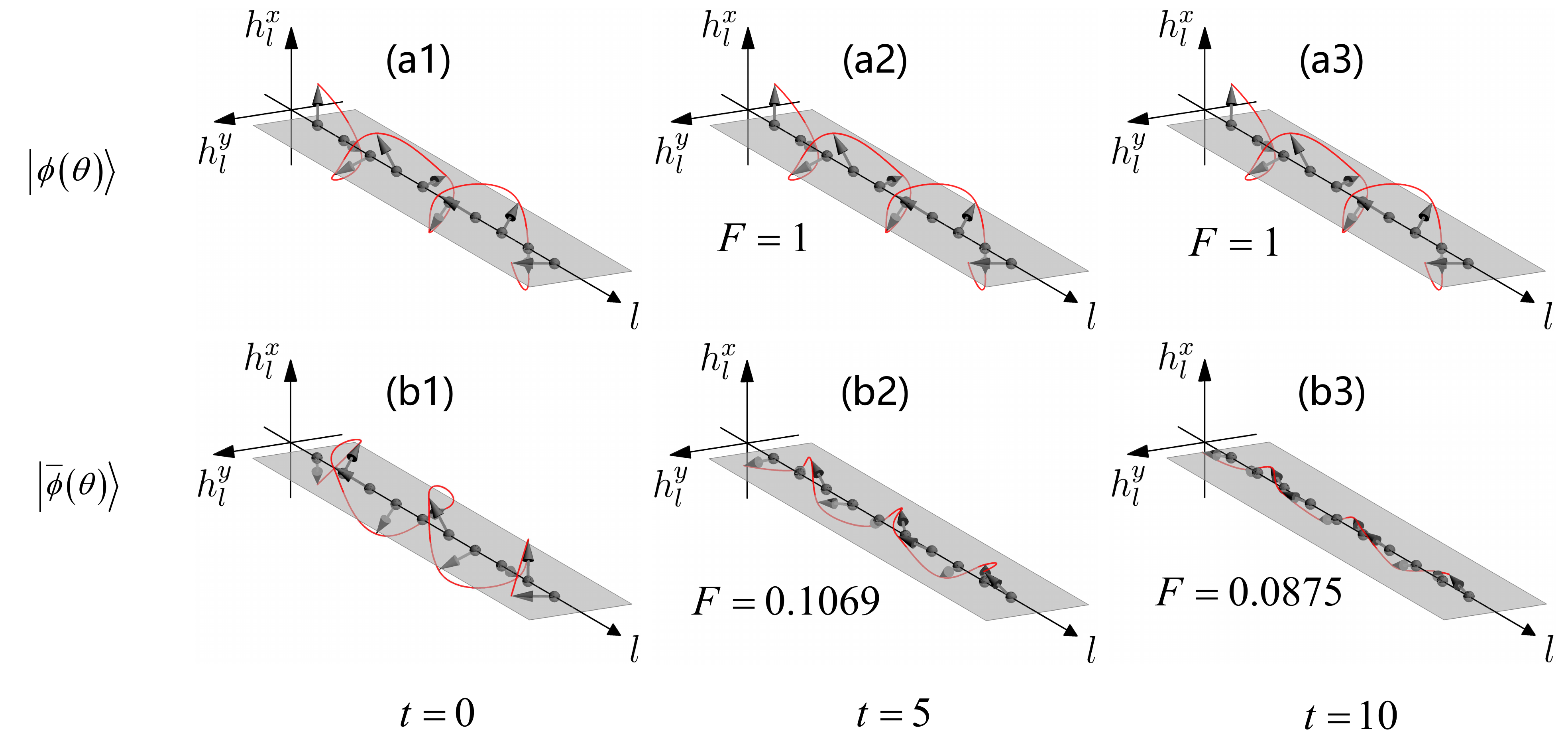}
\caption{Plots of the helix vector for evolved states defined in Eq. (%
\protect\ref{Psi_t}) at several representative instants for two types of
initial states . (a) $\left\vert \protect\phi \left( \protect\theta \right)
\right\rangle $\ and (b) $\left\vert \overline{\protect\phi }\left( \protect%
\theta \right) \right\rangle $. It shows that the evolved state for initial
state $\left\vert \protect\phi \left( \protect\theta \right) \right\rangle $%
\ is unchanged during the time evolution process while the one for $%
\left\vert \overline{\protect\phi }\left( \protect\theta \right)
\right\rangle $\ collapses from the initial state. The parameters are $%
\protect\theta =\protect\pi /3$, $q=6\protect\pi /10$, $\protect\lambda =1$\
and $N=10$. Fig. \protect\ref{fig2} plots the fidelity of the evolved state
for the initial state $\left\vert \overline{\protect\phi }\left( \protect%
\theta \right) \right\rangle $\ with different values of $\protect\lambda $.}
\label{fig1}
\end{figure*}

\section{Unidirectional helix states}

\label{Unidirectional helix states}

We begin with a Hamiltonian%
\begin{equation}
H=H_{0}+H_{\mathrm{DM}}  \label{H spin}
\end{equation}%
where $H_{0}$\ describes a quantum spin XXZ\ Heisenberg model

\begin{equation}
H_{0}=\sum_{j=1}^{N}[s_{j}^{x}s_{j+1}^{x}+s_{j}^{y}s_{j+1}^{y}+\cos
q(s_{j}^{z}s_{j+1}^{z}-\frac{1}{4})],
\end{equation}%
possessing the phantom spin helix states \cite{jepsen2022long}, and $H_{%
\mathrm{DM}}$\ is a DMI Hamiltonian 
\begin{equation}
H_{\mathrm{DM}}=i\frac{\lambda }{2}\sum_{j=1}^{N}\left( \bm{\tau }_{j}\times %
\bm{\tau }_{j+1}\right) \cdot \widehat{z},
\end{equation}%
with a unit vector $\widehat{z}=(0,0,1)$. Here $\mathbf{s}_{j}=\left(
s_{j}^{x},s_{j}^{y},s_{j}^{z}\right) $\ is the spin-$1/2$ operator, and
operator $\bm{\tau }_{j}=\left( \tau _{j}^{x},\tau _{j}^{y},\tau
_{j}^{z}\right) \ $is $p$-dependent and obtained from the unitary
transformation

\begin{eqnarray}
{\tau }_{j}^{x} &=&\cos \left( pj\right) s_{j}^{x}-\sin \left( pj\right)
s_{j}^{y},  \notag \\
{\tau }_{j}^{y} &=&\sin \left( pj\right) s_{j}^{x}+\cos \left( pj\right)
s_{j}^{y},  \notag \\
{\tau }_{j}^{z} &=&s_{j}^{z},
\end{eqnarray}%
which still satisfy the Lie algebra commutation relations 
\begin{equation}
\left[ \tau _{j}^{\mu },\tau _{j}^{\nu }\right] =2i\epsilon ^{\mu \nu
\lambda }\tau _{j}^{\lambda }.
\end{equation}%
The single-value condition of $\bm{\tau }_{j}$\ requires that $p=2\pi n/N$ ($%
n\in \left[ 0,N-1\right] $), while the periodic boundary condition of $H_{0}$%
\ has no restriction on the value of $q$. Nevertheless,\ this model becomes
special when we take $q=2\pi n/N$ ($n\in \left[ 0,N-1\right] $), and then
the Hamiltonian $H$ has translational symmetry.

For the self-consistency of the paper, we present a brief review on the
obtained result for $H_{0}$. There is a set of states of $H_{0}$, $\left\{
\left\vert \psi _{n}\right\rangle \right\} $ ($n\in \left[ 0,N\right] $),
which is defined as%
\begin{equation}
\left\vert \psi _{n}\right\rangle =\frac{1}{\Omega _{n}}\left( \tau
_{q}^{+}\right) ^{n}\left\vert \Downarrow \right\rangle ,
\end{equation}%
with the normalization factor $\Omega _{n}=\left( n!\right) \sqrt{C_{N}^{n}}$%
. Here the collective spin operator%
\begin{equation}
\tau _{q}^{+}=\sum_{j=1}^{N}\tau _{j}^{+}(q)=\sum_{j=1}^{N}e^{iqj}s_{j}^{+},
\end{equation}%
is essentially a spin magnon operator with wave vector $q$, which can excite
a magnon state when we apply it on the state $\left\vert \psi
_{0}\right\rangle =$ $\left\vert \Downarrow \right\rangle $ $%
=\prod_{j=1}^{N}\left\vert \downarrow \right\rangle _{j}$ and then we have 
\begin{equation}
\left\vert \psi _{N}\right\rangle =e^{iq\left( 1+N\right) N/2}\left\vert
\Uparrow \right\rangle  \label{SUS}
\end{equation}%
with $\left\vert \Uparrow \right\rangle =\prod\nolimits_{j=1}^{N}\left\vert
\uparrow \right\rangle _{j}$\textbf{. }Here, kets are defined as\textbf{\ }$%
s_{j}^{z}\left\vert \downarrow \right\rangle _{j}=-1/2\left\vert \downarrow
\right\rangle _{j}$\textbf{\ }and\textbf{\ }$s_{j}^{z}\left\vert \uparrow
\right\rangle _{j}=1/2\left\vert \uparrow \right\rangle _{j}$\textbf{.}
Importantly, straightforward derivation shows that $\left\{ \left\vert \psi
_{n}\right\rangle \right\} $\ is a set of eigenstates of $H_{0}$ with zero
energy, i.e.,%
\begin{equation}
H_{0}\left\vert \psi _{n}\right\rangle =0.
\end{equation}%
In parallel, another set of states $\left\{ \left\vert \overline{\psi }%
_{n}\right\rangle \right\} $ ($n\in \left[ 0,N\right] $), which is defined as%
\begin{equation}
\left\vert \overline{\psi }_{n}\right\rangle =\frac{1}{\Omega _{n}}\left(
\tau _{-q}^{+}\right) ^{n}\left\vert \Downarrow \right\rangle ,
\end{equation}%
are also eigenstates of $H_{0}$ with zero energy.\textbf{\ }We note\textbf{\ 
}that $\left\{ \left\vert \psi _{n}\right\rangle \right\} $ and $\left\{
\left\vert \overline{\psi }_{n}\right\rangle \right\} $\ share two common
states, i.e., $\left\vert \psi _{0}\right\rangle =\left\vert \overline{\psi }%
_{0}\right\rangle $ and $\left\vert \psi _{N}\right\rangle =$ $e^{iq\left(
1+N\right) N}\left\vert \overline{\psi }_{N}\right\rangle $.

Based on the two sets of states $\left\{ \left\vert \psi _{n}\right\rangle
\right\} $\ and $\left\{ \left\vert \overline{\psi }_{n}\right\rangle
\right\} $, two sets of helix states can be constructed, for instance, in
the form\ 
\begin{eqnarray}
&&\left\vert \phi (\theta )\right\rangle =\sum_{n}d_{n}\left\vert \psi
_{n}\right\rangle  \notag \\
&=&\overset{N}{\underset{j=1}{\prod }}[\cos \left( \theta /2\right)
\left\vert \downarrow \right\rangle _{j}-ie^{iqj}\sin \left( \theta
/2\right) \left\vert \uparrow \right\rangle _{j}],  \label{HS}
\end{eqnarray}%
where the coefficient%
\begin{equation}
d_{n}=\sqrt{C_{N}^{n}}\left( -i\right) ^{n}\sin ^{n}\left( \theta /2\right)
\cos ^{\left( N-n\right) }\left( \theta /2\right) .
\end{equation}%
We introduce a local vector $\mathbf{h}_{l}=\left(
h_{l}^{x},h_{l}^{y},h_{l}^{z}\right) $ with $h_{l}^{\alpha }=\left\langle
\psi \right\vert s_{l}^{\alpha }\left\vert \psi \right\rangle $ ($\alpha
=x,y,z$) to characterize the helicity of a given state $\left\vert \psi
\right\rangle $ at site $l$. The corresponding helix vector for $\left\vert
\phi (\theta )\right\rangle $ is 
\begin{equation}
\mathbf{h}_{l}=\frac{1}{2}[\sin \theta \sin \left( ql\right) ,\sin \theta
\cos \left( ql\right) ,-\cos \theta ],
\end{equation}%
which indicates that $\left\vert \phi (\theta )\right\rangle $\ is a helix
state for nonzero $\sin \theta $. Here $\theta $ is an arbitrary angle and
determines the profile of the state. In parallel state 
\begin{equation}
\left\vert \overline{\phi }(\theta )\right\rangle =\overset{N}{\underset{j=1}%
{\prod }}[\cos \left( \theta /2\right) \left\vert \downarrow \right\rangle
_{j}-ie^{-iqj}\sin \left( \theta /2\right) \left\vert \uparrow \right\rangle
_{j}],
\end{equation}%
is another one with opposite helicity. We note that both $\left\vert \phi
(\theta )\right\rangle $\ and $\left\vert \overline{\phi }(\theta
)\right\rangle $\ are also eigenstates of $H_{0}$\ with zero energy.

Now we switch on the term $H_{\mathrm{DM}}$ to see its effect,
which is the main point of this work. To investigate the action of $H_{%
\mathrm{DM}}$\ on the helix states $\left\vert \phi (\theta )\right\rangle $%
\ and $\left\vert \overline{\phi }(\theta )\right\rangle $, we rewrite it\
in the form%
\begin{eqnarray}
H_{\mathrm{DM}}(p) &=&i\lambda
\sum_{j=1}^{N}(e^{-ip}s_{j}^{+}s_{j+1}^{-}-e^{ip}s_{j}^{-}s_{j+1}^{+}) 
\notag \\
&=&i\lambda \sum_{j=1}^{N}s_{j}^{+}(e^{-ip}s_{j+1}^{-}-e^{ip}s_{j-1}^{-}).
\end{eqnarray}%
It can be checked that 
\begin{equation}
H_{\mathrm{DM}}(p)\left\vert \phi (\theta )\right\rangle =0,
\end{equation}%
when we take the resonant condition $p=q$. In fact, the commutation
relations \textbf{\ }%
\begin{equation}
\left[ H_{\mathrm{DM}}(q),\tau _{q}^{+}\right] =i\lambda
\sum_{j=1}^{N}e^{iq}s_{j}^{+}(s_{j-1}^{z}-s_{j+1}^{z}),
\end{equation}%
and%
\begin{equation}
\left[ \left[ H_{\mathrm{DM}}(q),\tau _{q}^{+}\right] ,\tau _{q}^{+}\right]
=0,
\end{equation}%
result in%
\begin{equation}
H_{\mathrm{DM}}(q)\left\vert \psi _{n}\right\rangle =0,
\end{equation}%
for any $n$. This means that the helix states $\left\vert \phi (\theta )\right\rangle $
are also the eigenstates of $H_{0}+H_{\mathrm{DM}}(q)$\ with zero energy.
However, $\left\vert \overline{\phi }(\theta )\right\rangle $\ is the
eigenstate of $H_{0}+H_{\mathrm{DM}}(-q)$, but not of $H_{0}+H_{\mathrm{DM}%
}(q)$. Put differently, $\left\vert \phi (\theta )\right\rangle $\ does not
undergo any evolution under the $H_{0}+H_{\mathrm{DM}}(q)$, whereas will
deviate from $\left\vert \phi (\theta )\right\rangle $\ driven by $H_{0}+H_{%
\mathrm{DM}}(-q)$. The DMI $H_{\mathrm{DM}}(q)$\ can select its resonant
helix state from the others. To demonstrate and verify our result, we
consider the time evolution process driven by the Hamiltonian $H_0+H_{\mathrm{DM}}(q)$. We employ the fidelity

\begin{equation}
F\left( t\right) =\left\vert \langle \Psi (0)\left\vert \Psi
(t)\right\rangle \right\vert ^{2}  \label{fidelity}
\end{equation}%
and the helix vector\ $\mathbf{h}_{l}(t)$\ to characterize the evolved state 
\begin{equation}
\left\vert \Psi (t)\right\rangle =\exp \left[ -iH_{0}t-iH_{\mathrm{DM}}(q)t%
\right] \left\vert \Psi (0)\right\rangle .  \label{Psi_t}
\end{equation}
Here two initial states $\left\vert \Psi (0)\right\rangle =\left\vert \phi
(\theta )\right\rangle $\ and $\left\vert \overline{\phi }(\theta
)\right\rangle $\ are considered.\ The plots of$\ \mathbf{h}_{l}(t)\ $and
the corresponding$\ F\left( t\right) $\ in Fig. \ref{fig1} and Fig. \ref%
{fig2}\ for the finite system show that $\left\vert \Psi (t)\right\rangle $\
remains unchanged for $\left\vert \phi (\theta )\right\rangle $, while\
another initial state $\left\vert \overline{\phi }(\theta )\right\rangle $
collapses from a helix state as expected. Then we conclude that the
Hamiltonian $H_{0}+H_{\mathrm{DM}}(q)$ can only support the unidirectional
helix states with the wave vector $q$.

\begin{figure}[t]
\centering\includegraphics[width=0.45\textwidth]{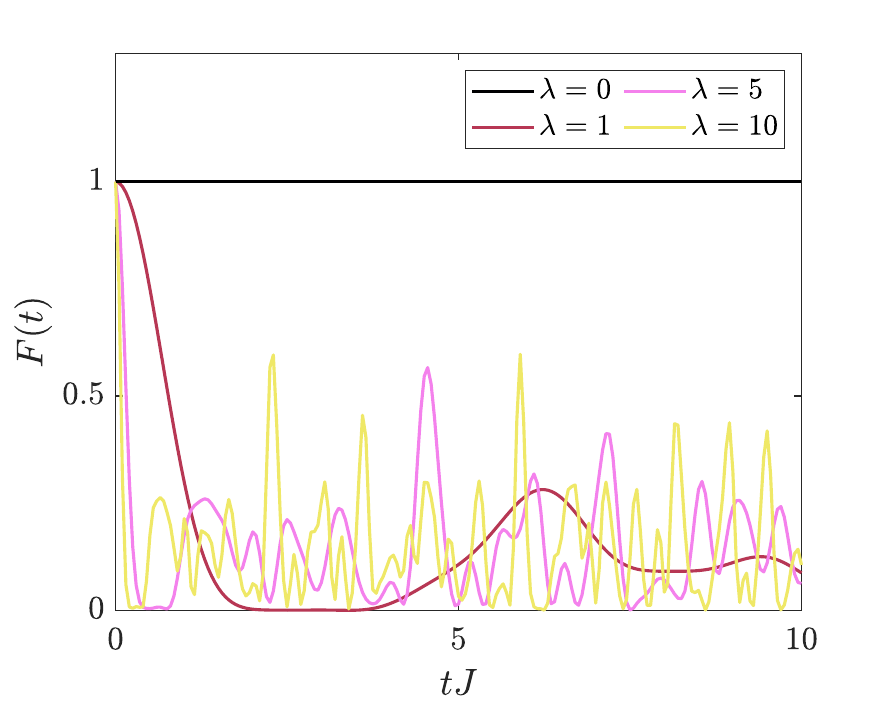} 
\caption{Plot of the fidelity defined in Eq. (\protect\ref{fidelity}) for the time evolution of initial state $\left\vert \overline{\protect\phi }\left( \protect\theta \right) \right\rangle $ under the
Hamiltonian $H$ with different values of $\protect\lambda $. Other parameters are $\protect\theta =\protect\pi /3$, $q=6\protect\pi /10$, $N=10$. We find that (i) the
fidelity remains unity for case with zero $\protect\lambda $; (ii)
when $\protect\lambda $ is nonzero, the fidelity decays and does
not revival, which accords with our analytical prediction. The decay rate
increases as $\protect\lambda $ increases.} \label{fig2}
\end{figure}

\begin{figure*}[t]
\centering\includegraphics[width=0.9\textwidth]{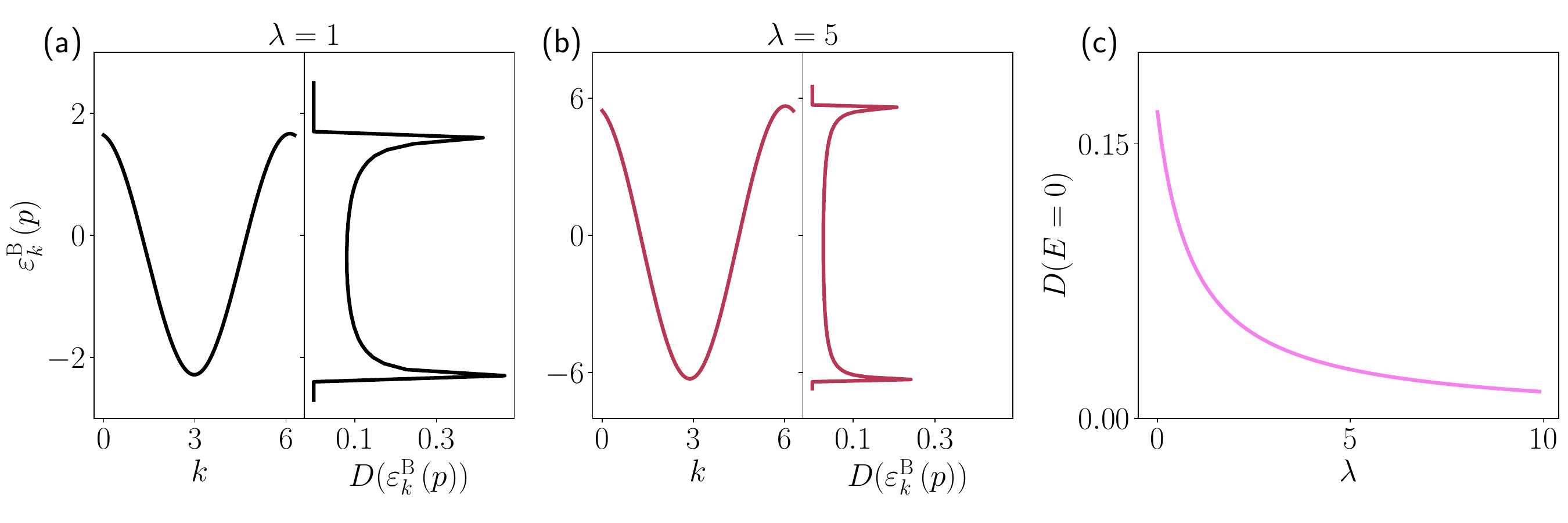}
\caption{(a) and (b) The energy levels $\protect\varepsilon _{k}^{B}\left(
p\right) $ as a function of $k$ (left side) and the DOS $D\left( \protect%
\varepsilon _{k}^{B}\left( p\right) \right) $ as a function of $\protect%
\varepsilon _{k}^{B}\left( p\right) $ (right side) with $\protect\lambda =1$
and $\protect\lambda =5$, respectively. (c) The DOS at $\protect\varepsilon %
_{k}^{B}\left( p\right) =0$ as a function of $\protect\lambda $. It
shows that the DMI can shift the zero energy levels of all $\left\{
\left\vert \overline{\protect\psi }_{n}\right\rangle \right\} $, except $%
\left\vert \overline{\protect\psi }_{0}\right\rangle $ and $\left\vert 
\overline{\protect\psi }_{N}\right\rangle $, as well as other energy levels
close to zero, away from zero energy. Other parameters are $p=2\protect\pi %
/5 $. Other parameter is $N=20000$.}
\label{fig3}
\end{figure*}

\section{Bosonic representation}

\label{Bosonic representation}

In order to understand the obtained results in last section, especially the
influence of $H_{\mathrm{DM}}(q)$\ on the other set of states $\left\{
\left\vert \overline{\psi }_{n}\right\rangle \right\} $, we simplify the
Hamiltonian based on the approximation method. We will show that the DMI can
not only single out the unidirectional helix state, but also separate it
from nearby energy levels around zero energy.

In this section, we investigate our results in an approximate way, which
reveals the underlying mechanism of the helix state and predicts its
robustness attributed to the DMI. We introduce the HP-transformation%
\begin{eqnarray}
s_{j}^{+} &=&\left( s_{j}^{-}\right) ^{\dag }=\sqrt{1-b_{j}^{\dagger }b_{j}}%
b_{j},  \notag \\
s_{j}^{z} &=&\frac{1}{2}-b_{j}^{\dagger }b_{j},
\end{eqnarray}%
to express the spin operators in terms of boson creation and annihilation
operators, $b_{j}$ and $b_{j}^{\dagger }$, where satisfying 
\begin{equation}
\left[ b_{j},b_{l}^{\dagger }\right] =\delta _{jl},\left[ b_{j},b_{l}\right]
=0.
\end{equation}%
Taking the approximation $\sqrt{1-b_{j}^{\dagger }b_{j}}\simeq 1$ in the
subspace $\frac{1}{N}\left\langle \sum_{j=1}^{N}b_{j}^{\dag
}b_{j}\right\rangle \ll 1$, the Hamiltonian $H$ can be mapped into an
effective on in the form%
\begin{equation}
H^{\mathrm{B}}=H_{0}^{\mathrm{B}}+H_{\mathrm{DM}}^{\mathrm{B}},
\end{equation}%
with%
\begin{equation}
H_{0}^{\mathrm{B}}=\frac{1}{2}\sum_{j=1}^{N}(b_{j}^{\dagger
}b_{j+1}+b_{j+1}^{\dagger }b_{j}-2\cos qb_{j}^{\dag }b_{j}),
\end{equation}%
and%
\begin{equation}
H_{\mathrm{DM}}^{\mathrm{B}}=i\frac{\lambda }{2}\sum \left(
e^{-ip}b_{j}^{\dagger }b_{j+1}-e^{ip}b_{j+1}^{\dagger }b_{j}\right) .
\end{equation}%
Obviously, $H_{0}^{\mathrm{B}}$\ is a simple tight-binding ring for
noninteraction bosons, while $H_{\mathrm{DM}}^{\mathrm{B}}$ is the ring with
a magnetic flux threading through it. Both of them are exactly solvable,
providing a way to understand the original Hamiltonian.

Applying the Fourier transformation%
\begin{equation}
b_{k}=\frac{1}{\sqrt{N}}\sum_{j}e^{-ikj}b_{j},
\end{equation}%
the Hamiltonian becomes diagonal%
\begin{equation}
H^{\mathrm{B}}=\sum_{k\in \text{BZ}}\varepsilon _{k}^{\mathrm{B}%
}b_{k}^{\dagger }b_{k},
\end{equation}%
with the spectrum%
\begin{equation}
\varepsilon _{k}^{\mathrm{B}}\left( p\right) =\cos k-\cos q-\lambda \sin
\left( k-p\right) .
\end{equation}%
In parallel, we can construct the many-body condensate states

\begin{eqnarray}
\left\vert \psi _{n}\right\rangle &=&\frac{1}{\Omega _{n}}\left( b_{q}^{\dag
}\right) ^{n}\left\vert 0\right\rangle ,  \notag \\
\left\vert \overline{\psi }_{n}\right\rangle &=&\frac{1}{\Omega _{n}}\left(
b_{-q}^{\dag }\right) ^{n}\left\vert 0\right\rangle ,
\end{eqnarray}%
which are eigenstates of $H_{0}^{\mathrm{B}}$\ with zero energy. The states $%
\left\vert \psi _{n}\right\rangle $ are also the eigenstates of $H_{\mathrm{%
DM}}^{\mathrm{B}}$ with zero energy when taking $k=p=q$, i.e., $\varepsilon
_{q}^{\mathrm{B}}\left( q\right) =0$. In contrast, we have $\varepsilon
_{-q}^{\mathrm{B}}\left( q\right) =\lambda \sin \left( 2q\right) \neq 0$%
\textbf{\ }when taking $k=-p=-q$.

Thus far, all the conclusions from the spin model $H$ hold for the boson
Hamiltonian $H^{\mathrm{B}}$. On the other hand, it is clear that when $H_{%
\mathrm{DM}}^{\mathrm{B}}$ is applied to $\left\vert \overline{\psi }%
_{n}\right\rangle $ or the other eigenstates around the zero engergy level,
an extra nonzero energy is acquired. In this sense, the DMI can isolate the
zero energy levels for $\left\{ \left\vert \psi _{n}\right\rangle \right\} $
by shifting the zero energy levels of $\left\{ \left\vert \overline{\psi }%
_{n}\right\rangle \right\} $\ (except $\left\vert \overline{\psi }%
_{0}\right\rangle $ and $\left\vert \overline{\psi }_{N}\right\rangle $) and
other near-zero energy levels from zero energy.\ This can be demonstrated
from the density of state (DOS) of the system near zero energy. In general,
the DOS at energy level $E$ is defined as%
\begin{equation}
D\left( E\right) =\frac{1}{N}\left\vert \frac{\text{d}N\left( E\right) }{%
\text{d}E}\right\vert ,  \label{dos}
\end{equation}%
where d$N\left( E\right) $\ indicates the number of energy levels that
appear in the interval $\left[ E,E+\text{d}E\right] $, with the
normalization factors $1/N$. For the present boson system, the
single-particle DOS can be obtained in the form%
\begin{equation}
D\left( E\right) =\frac{1}{2\pi \sqrt{\lambda ^{2}+1+2\lambda \sin p-E^{2}}},
\end{equation}%
from the spectrum $\varepsilon _{k}^{\mathrm{B}}\left( q\right) $. In Fig. %
\ref{fig3} (a) and (b), we plot the energy levels $\varepsilon _{k}^{\mathrm{%
B}}\left( p\right) $\ as a function of momentum $k$ (left side) \ and DOS as
a function of $\varepsilon _{k}^{\mathrm{B}}\left( p\right) $\ (right side)\
for the systems with $\lambda =1$\ and $\lambda =5$, respectively. Fig. \ref%
{fig3} (c) shows that the DOS near the zero energy becomes relatively low as 
$\lambda $\ increases.

\begin{figure*}[t]
\centering\includegraphics[width=0.9\textwidth]{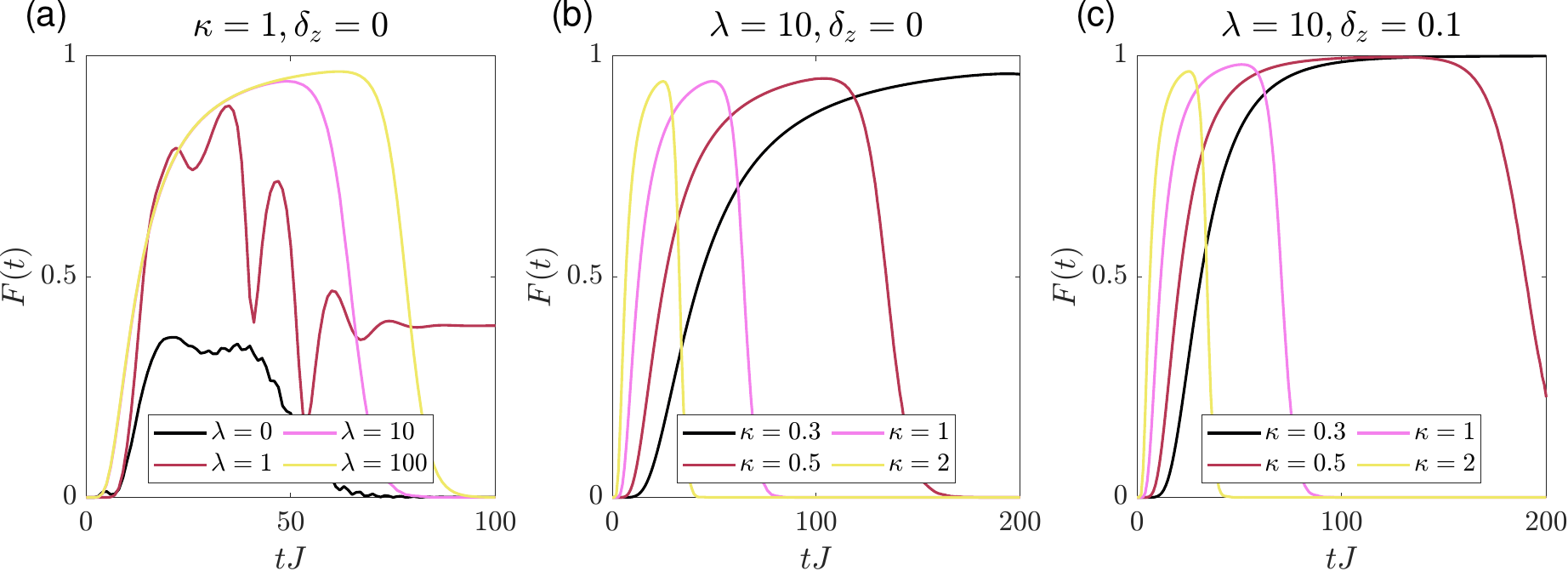}
\caption{The fidelity of $\left\vert \widetilde{\protect\psi }\left(
t\right) \right\rangle $ with $|\widetilde{\protect\psi }_{N}\rangle $
driven by the Hamiltonian $H_{\mathrm{dri}}$ with different $\protect\kappa $
and $\protect\lambda .$ (a) and (b) reveal that the time evolution of
fidelity exhibits the behavior of exceptional point dynamics under large $%
\protect\lambda $ and small $\protect\kappa $ conditions. By adding
a compensation term $i\protect\delta \widetilde{s}_{l}^{z}(q)$, the
fidelity can be improved to reach unity, as demonstrated by (c).}
\label{fig4}
\end{figure*}

\begin{figure}[t]
\centering\includegraphics[width=0.4\textwidth]{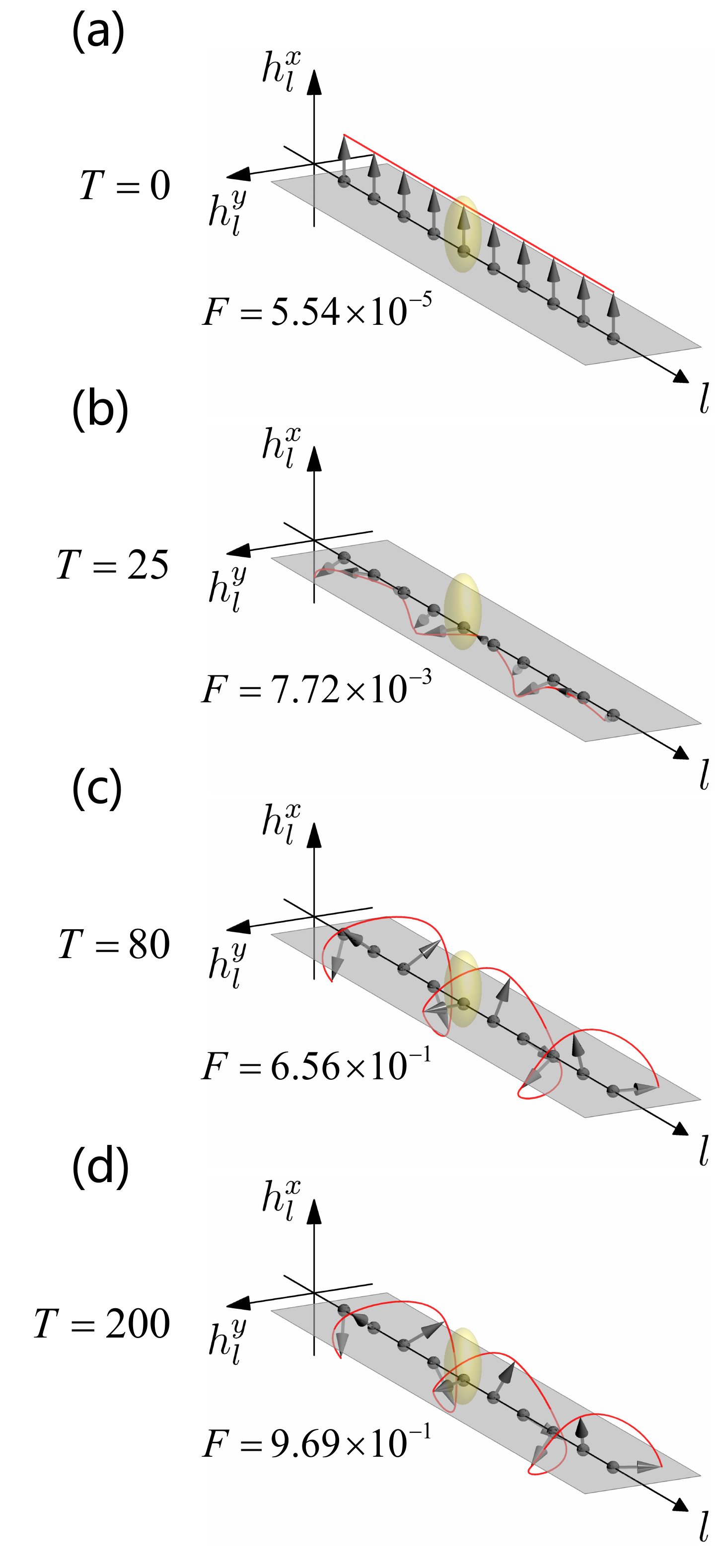} 
\caption{The plots of the helix vector for several time evolved states
driven by the Hamiltonian defined in Eq. (\protect\ref{DEVN}). The
Hamiltonian is modified by adding a complex local field at the site
$j=5$, represented by the yellow ellipsoidal area. The initial state,
prepared according to Eq. (\protect\ref{SUS}), evolves over time to yield
the phantom helix state defined in Eq. (\protect\ref{PWW}), as demonstrated
by the plots. Other parameters are $\protect\theta =\protect\pi /3$, $q=6\protect\pi /10$, $\protect\lambda =10$, $\protect\kappa =0.3$, $\protect\delta =0.1$ and $N=10$.}
\label{fig5}
\end{figure}

\section{Dynamic demonstration}

\label{Dynamic demonstration}

Now, we turn to the application of the above results. We will show that a
unidirectional helix state can be generated by an external impact that is
described by a\ non-Hermitian term $H_{\mathrm{I}}$.\ It can be a local
resonant perturbation.\ We start with the investigation for an exactly
solvable case, in which $H_{\mathrm{I}}$ arises from a complex
spatially modulated field $\mathbf{B}_{j}=(B_{j}^{x},B_{j}^{y},B_{j}^{z})$,
i.e.,%
\begin{equation}
H_{\mathrm{I}}=\sum_{j=1}^{N}\mathbf{B}_{j}\cdot \mathbf{s}_{j},  \label{HI}
\end{equation}%
with the reduced components in the unit of strength $\kappa $,%
\begin{eqnarray}
B_{j}^{x}/\kappa &=&\cos ^{2}\left( \theta /2\right) +e^{-i2qj}\sin
^{2}\left( \theta /2\right) ,  \notag \\
B_{j}^{y}/\kappa &=&\cos ^{2}\left( \theta /2\right) -ie^{-i2qj}\sin
^{2}\left( \theta /2\right) ,  \notag \\
B_{j}^{z}/\kappa &=&ie^{-iqj}\sin \theta .
\end{eqnarray}%
It is a little complicated at first glance. However, it only contains the
information encoded by two parameters\ ($\theta ,q$). In the following, we
will show that such a deliberately designed field can be utilized to
dynamically generate the steady helix state $\left\vert \phi (\theta
)\right\rangle $ from a trivial state, and importantly, such a task can also
be accomplished\ by a local field.

The field $\mathbf{B}_{j}$\ is taken based on the helix state $\left\vert
\phi (\theta )\right\rangle $\ in Eq. (\ref{HS}), which can be rewritten in
the form%
\begin{equation}
\left\vert \phi (\theta )\right\rangle =\prod_{j}^{N}|\widetilde{\downarrow }%
\rangle _{j},
\end{equation}%
with%
\begin{equation}
|\widetilde{\downarrow }\rangle _{j}=\cos \left( \theta /2\right) \left\vert
\downarrow \right\rangle _{j}-ie^{iqj}\sin \left( \theta /2\right)
\left\vert \uparrow \right\rangle _{j}.
\end{equation}%
Obviously, state $|\widetilde{\downarrow }\rangle _{j}$ can be obtained by a
rotation from $\left\vert \downarrow \right\rangle _{j}$. Accordingly, a
related set of spin operators can be constructed as%
\begin{eqnarray}
&&\widetilde{s}_{j}^{+}(q)=\left[ \widetilde{s}_{j}^{-}(q)\right] ^{\dag
}=e^{-iqj}\times \\
&&\left[ e^{iqj}\cos ^{2}\left( \theta /2\right) s_{j}^{+}+e^{-iqj}\sin
^{2}\left( \theta /2\right) s_{j}^{-}+i\sin \theta s_{j}^{z}\right] ,  \notag
\end{eqnarray}%
and%
\begin{equation}
\widetilde{s}_{j}^{z}(q)=\frac{1}{2}\left[ \widetilde{s}_{j}^{+}(q),%
\widetilde{s}_{j}^{-}(q)\right] ,
\end{equation}%
satisfying $\widetilde{s}_{j}^{z}(q)|\widetilde{\downarrow }\rangle _{j}=-%
\frac{1}{2}|\widetilde{\downarrow }\rangle _{j}$. Introducing the collective
operators%
\begin{eqnarray}
\widetilde{s}_{q}^{+} &=&\left( \widetilde{s}_{q}^{-}\right) ^{\dag
}=\sum_{j=1}^{N}\widetilde{s}_{j}^{-}(q),  \notag \\
\widetilde{s}_{q}^{z} &=&\sum_{j=1}^{N}\widetilde{s}_{j}^{z}(q),
\end{eqnarray}%
and defining a tilted ferromagnetic state%
\begin{equation}
|\widetilde{\psi }_{0}\rangle =\left\vert \phi (\theta )\right\rangle
=\prod_{j}^{N}|\widetilde{\downarrow }\rangle _{j},
\end{equation}%
then the Hamiltonian we defined in Eq. (\ref{HI}) can be expressed as%
\begin{equation}
H_{\mathrm{I}}=\kappa \widetilde{s}_{q}^{+},
\end{equation}%
and%
\begin{eqnarray}
\widetilde{s}_{q}^{z}|\widetilde{\psi }_{0}\rangle &=&-\frac{N}{2}|%
\widetilde{\psi }_{0}\rangle ,  \notag \\
H_{\mathrm{I}}^{\dag }|\widetilde{\psi }_{0}\rangle &=&0,
\end{eqnarray}%
which connects the perturbation $H_{\mathrm{I}}$\ with the helix state. In
parallel, one can also use the operator $\widetilde{s}_{q}^{+}$\ to
construct a series of the state%
\begin{equation}
|\widetilde{\psi }_{n}\rangle =\frac{1}{\Omega _{n}}\left( \widetilde{s}%
_{q}^{+}\right) ^{n}|\widetilde{\psi }_{0}\rangle ,
\end{equation}%
with $n\in \left[ 0,N-1\right] $. We note that state%
\begin{eqnarray}
&&\left\vert \widetilde{\psi }_{N}\right\rangle =\prod_{j}^{N}\left\vert 
\widetilde{\uparrow }\right\rangle _{j}  \notag \\
&=&\prod_{j}^{N}\left[ \cos \left( \theta /2\right) \left\vert \uparrow
\right\rangle _{j}-ie^{-iqj}\sin \left( \theta /2\right) \left\vert
\downarrow \right\rangle _{j}\right] ,  \label{PWW}
\end{eqnarray}%
is another tilted ferromagnetic state. The relation\textbf{\ }%
\begin{equation}
\widetilde{s}_{q}^{+}\left\vert \widetilde{\psi }_{n}\right\rangle =\sqrt{%
\frac{\left( N-n\right) }{\left( n+1\right) }}\left\vert \widetilde{\psi }%
_{n+1}\right\rangle ,  \label{UPW}
\end{equation}%
ensures the existence of an invariant ($N+1$)-D\ subspace for $H_{\mathrm{I}%
} $, spanned by the set of states $\left\{ |\widetilde{\psi }_{n}\rangle
\right\} $.

Furthermore, the matrix representation of the Hamiltonian $H_{\mathrm{I}}$
in the ($N+1$)-D\ subspace is an $\left( N+1\right) \times \left( N+1\right) 
$ non-Hermitian matrix $M$ with nonzero matrix elements

\begin{equation}
\left( M\right) _{n+2,n+1}\mathbf{=}\kappa \sqrt{\frac{\left( N-n\right) }{%
\left( n+1\right) }}\mathbf{.}
\end{equation}%
It is obviously $M$\ is a nilpotent matrix, i.e.%
\begin{equation}
\left( M\right) ^{N+1}=0,
\end{equation}%
or, in other words, $M$\textbf{\ }is\textbf{\ }an $N+1$-order Jordan block.
This means that $H$ has an $N+1$-order\ EP with coalescing state is $|%
\widetilde{\psi }_{N}\rangle $, i.e., there are $N+1$ energy levels that
share a common eigenstate $|\widetilde{\psi }_{N}\rangle $. The dynamics for
any initial state $\left\vert \Psi (0)\right\rangle $ in this subspace is
governed by the time evolution operator%
\begin{equation}
U(t)=e^{-iMt}=\sum_{l=0}^{N}\frac{1}{l!}\left( -iMt\right) ^{l}.
\end{equation}%
Then the evolved state becomes a steady state and approches 
\begin{equation}
\lim_{t\rightarrow \infty }\frac{U(t)\left\vert \Psi (0)\right\rangle }{%
\left\vert U(t)\left\vert \Psi (0)\right\rangle \right\vert }=|\widetilde{%
\psi }_{N}\rangle ,
\end{equation}%
after a sufficiently long time.

Now, we consider a similar procedure for the case with a simple local
perturbation term%
\begin{equation}
H_{\mathrm{I}}=\kappa \widetilde{s}_{l}^{+}(q).
\end{equation}%
We note that we still have%
\begin{equation}
\widetilde{s}_{l}^{+}(q)|\widetilde{\psi }_{N}\rangle =0,
\end{equation}%
for the local operator. According to the theorem in Ref. \cite{WPPRB}, the
state $|\widetilde{\psi }_{N}\rangle $\ is still the coalescing state for $%
H_{\mathrm{I}}$. Accordingly, the matrix representation of Hamiltonian $H_{%
\mathrm{I}}$ is an $\left( N+1\right) \times \left( N+1\right) $ matrix $%
M^{\prime }$ with 
\begin{equation}
M^{\prime }=\frac{1}{N}M.
\end{equation}%
Two matrix representations $M$ and $M^{\prime }$ are different, the former
is exact, while the latter is approximate. However, the term $H_{\mathrm{DM}%
} $ enhances the validity\ of the approximation\ on $M^{\prime }$ by
shrinking the dimension of the degenerate subspace. In the following, we
investigate the dynamics driven by the Hamiltonian%
\begin{equation}
H_{\mathrm{DRVN}}=H+H_{\mathrm{I}}.  \label{DEVN}
\end{equation}

The normalized evolved state has the form%
\begin{equation}
\left\vert \Psi \left( t\right) \right\rangle =\frac{\exp \left( -iH_{%
\mathrm{DRVN}}t\right) \left\vert \Psi (0)\right\rangle }{\left\vert \exp
\left( -iH_{\mathrm{DRVN}}t\right) \left\vert \Psi (0)\right\rangle
\right\vert },
\end{equation}%
where the initial state $\left\vert \Psi (0)\right\rangle $\textbf{\ }can be
any combination of the set of states $\left\{ |\widetilde{\psi }_{n}\rangle
\right\} $, that means 
\begin{equation}
\left\vert \Psi (0)\right\rangle =\sum_{n=0}^{N}a_{n}|\widetilde{\psi }%
_{N}\rangle .  \label{AC}
\end{equation}%
The fidelity of the evolved state driven from the initial state\textbf{\ }$%
\left\vert \Psi (0)\right\rangle $ with the target state $|\widetilde{\psi }%
_{N}\rangle $ 
\begin{equation}
F(t)=\left\vert \langle \widetilde{\psi }_{N}\left\vert \Psi \left( t\right)
\right\rangle \right\vert ^{2},
\end{equation}%
measures the efficiency of the scheme. Here we consider the Dirac basis \cite{LSPRA,YCTPRR,SYBarxiv}. Numerical simulation is performed by
taking $a_{n}=\delta _{n,0}$ for the initial state, which is a simply
ferromagnetic state. The results for different parameters are plotted in
Fig. \ref{fig4}. Fig. \ref{fig4}(a) shows that the term $H_{\mathrm{DM}}$\
enhances the validity\ of the approximation\ on $M^{\prime }$, providing a
near perfect Jordan block showing evident EP dynamics. Fig. \ref{fig4}(b)
shows that the value of $\kappa $\ has no obvious influence on the peak of%
\textbf{\ }$F(t).$ They reveal that the time evolution of fidelity exhibits
the behavior of exceptional point dynamics under large $\lambda $ and small $%
\kappa $ conditions.

However, we also find that the maximal $F(t)$\ cannot reaches to unity and
decays quickly after a long period of time. This is because the subspace $%
\left\{ |\widetilde{\psi }_{n}\rangle \right\} $\ is the perfect invariant
subspace for $\widetilde{s}_{q}^{+}$, but not the perfect one for $%
\widetilde{s}_{l}^{+}(q)$. The evolved state may leak out of the subspace
and collapse to the eigenstate of $H_{\mathrm{DRVN}}$\ with maximal positive
imaginary part of energy. To avoid this, one can add a compensation term $%
i\delta \widetilde{s}_{l}^{z}(q)$\ ($\delta >0$), i.e., taking%
\begin{equation}
H_{\mathrm{I}}=\kappa \widetilde{s}_{l}^{+}(q)+i\delta \widetilde{s}%
_{l}^{z}(q),
\end{equation}%
to suppress leakage. The results for nonzero $\delta $ are plotted in Fig. %
\ref{fig4} (c). Fig. \ref{fig4}(c) shows that $F(t)$\ can reach unity.
Another dynamic process is plotted in Fig. \ref{fig5}. Fig. \ref{fig5}
illustrates the helix vector for several involved states and the
corresponding $F(t)$,\ driven by the Hamiltonian $H_{\mathrm{DRVN}}$. It
demonstrates that any initial state has the form we defined in Eq. (\ref{AC}%
) evolves over time to yield the phantom helix state defined in Eq. (\ref%
{PWW}). Specifically, we choose the $\left\vert \psi _{N}\right\rangle $ we
have defined in Eq. (\ref{SUS}). Note that $\kappa \widetilde{s}%
_{l}^{+}(q)+i\delta \widetilde{s}_{l}^{z}(q)$\ is also $\theta -$dependent,
and then the target state $|\widetilde{\psi }_{N}\rangle $ can be set for
arbitrary $\theta $. Then one can dynamically generate a perfect
unidirectional helix state by a single non-Hermitian impurity, which
contains the information of the target helix state\ ($\theta ,q$) in Eq. (%
\ref{PWW}). Finally, we would like to point out that $H_{\mathrm{I}}$\ and
the initial state are equal for both the target state $|\widetilde{\psi }%
_{N}\rangle $\ and its counterpart helix state with opposite direction in
the absence of DMI.\ Then the final state should not be a unidirectional
helix state.

\section{Summary}

\label{Summary}

In summary, we have studied the influence of the DMI on bidirectionally
equal phantom helix states in the XXZ Heisenberg model. Based on the
analysis by HP transformation, we show that only the helix states in one
direction are still eigenstates with zero energy in the presence of the
resonant DMI, while those in the opposite direction are left out of the
subspace. Then the DMI can act as a filter for unidirectional helix states.
This allows the formation of a Jordan block in the matrix representation of
the Hamiltonian when the non-Hermitian term is added, in which the
coalescing state can be the unidirectional helix states on demand. We
numerically demonstrate the dynamic generation processes of unidirectional
phantom helix states stimulated by a local non-Hermitian impurity. Our
findings provide a way to select a unidirectional helix state and offer a
method for the efficient stimulation of a spin helix state by a local
non-Hermitian impurity via the quench dynamic process. It is expected to be
insightful for the investigation of helix states in experiments.

This work was supported by National Natural Science Foundation of China
(under Grant No. 11874225).

\end{document}